\documentclass[superscriptaddress]{revtex4-1}

\usepackage{subfigure,epsfig,algorithm2e}

\begin{document}

\title{Automatic City Region Analysis for Urban Routing}

\author{Kai Zhao}%
\email{kai.zhao@nyu.edu}
\affiliation{Department of Computer Science, University of Helsinki, Helsinki, Finland}
\affiliation{CUSP, New York University, New York, USA}

\author{Mohan Prasath C}%
\affiliation{Department of Computer Science, University of Helsinki, Helsinki, Finland}

\author{Sasu Tarkoma}%
\email{sasu.tarkoma@cs.helsinki.fi}
\affiliation{Department of Computer Science, University of Helsinki, Helsinki, Finland}

\begin{abstract}

There are different functional regions in cities such as tourist attractions, shopping centers, workplaces and residential places. 
The human mobility patterns for different functional regions are different, e.g., people usually go to work during daytime on weekdays, and visit shopping centers after work. 
In this paper, we analyse urban human mobility patterns and infer the functions of the regions in three cities. The analysis is based on three large taxi GPS datasets in Rome, San Francisco and Beijing containing 21 million, 11 million and 17 million GPS points respectively. 
We categorized the city regions into four kinds 
of places, workplaces, entertainment places, residential places and other places. 
First, we provide a new quad-tree region division method based on the taxi visits. 
Second, we use the association rule to infer the functional regions in these three cities according to temporal human mobility patterns. 
Third, we show that these identified functional regions can help us deliver data in network applications, such as urban Delay Tolerant Networks (DTNs), more efficiently. 
The new functional-regions-based DTNs algorithm
achieves up to 183\% improvement in terms
of delivery ratio. \footnote{This paper was accepted by ICDM workshop, in Atlantic City, NJ, 2015}

\end{abstract}

\maketitle 

\section{Introduction}

Nowadays, a variety of urban human mobility data have been gathered and published 
\cite{Beijing1, Beijing2, SanF, Rome, SR15, MobiSys15}. The 
pervasive GPS data can be collected by mobile phones or car GPS navigation devices. A mobile operator can track people's 
movement in cities based on their cellular network location. 
This urban human mobility data contains rich knowledge about location information and can help in addressing many urban challenges, such as air pollution problems \cite{Uair}. 
Urban human mobility patterns pertain to how people move in cities, for example, driving home, walking to work or utilizing public transportation to shopping 
centers. 

This paper aims to utilize the 
knowledge of urban human mobility patterns to identify the 
functional regions of three cities, Rome, San Francisco and Beijing. 
A functional region is a region (we use grid here) that has a specific characteristic such as tourist attractions, shopping centers, educational areas, workplaces or residential places. 
The human mobility patterns for different function regions are different \cite{siyuan1,siyuan2}. 
People usually go to work on weekdays during daytime, visit the 
entertainment places such as shopping centers after work, and stay at home during the night. 
Such temporal human mobility patterns can help us to identify the functional regions of a city. 

In this paper, we categorize the city regions into four kinds 
of places, workplaces, entertainment places, residential places and other places. Our analysis is based on three taxi GPS datasets in three cities, Rome, San Francisco and Beijing containing 21 million, 11 million and 17 million GPS points respectively. 
We infer the temporal human mobility of these three cities based on the taxi trips. 
Since taxis carry people to different places at different times, taxi trips reflect how people move in a city. We define a taxi's trip as two GPS points from picking up the passenger until dropping off the 
passenger. The taxi trips reflect the actual human movement inside the city.

First, to divide a city into smaller regions, we provide a new quad-tree-based region division method based on taxi visits. 
In the previous research, Yuan et al. divide a city according to the road networks 
\cite{regionfunction}. 
We find that the road networks are highly correlated with the taxi visits by the Pearson co-efficiency test. 
We can divide the city into different regions only based on the taxi visits without the road network information that can be difficult to obtain. 
The reason behind the positive correlation between road networks and taxi visits 
is that the taxi visits are also positively correlated with the population density \cite{ExponentialOne}. When there are 
more people living in a 
region, there are more roads in that region providing connections to other regions and more taxis picking up/dropping off passengers. 

The population of a city is exponentially distributed, decreasing from the 
city center to rural areas \cite{ExponentialOne}. A quad-tree-based region division method 
can capture the details of the human mobility in the city center, while reducing unnecessary 
calculation pertaining to sparse rural areas.

Second, we use the association rule to infer the functions of different regions in these three cities according to the temporal travel patterns. We construct a boolean table for each taxi's 
visits in each region and obtain the frequent item-sets (most dominant visits) at different times. 
The function of a region is constructed 
based on the most dominant visits of that region. We define workplaces where people usually visit during the day-time on weekdays, 
entertainment places where people usually stay during the evening on weekdays or during 
weekends, 
residential places where people usually stay overnight, and other places with no identical temporal mobility patterns. 

Note that, a region can have multiple functions, i.e., a regions can be a place of work, residence and entertainment. We define the single function of a region only based on the period that most dominant human visits happened, e.g., a work-place where most people come to work instead of living there. Most of the regions have a specific function. For example, there are some people living in Manhattan in New York. However, most of the people come to Manhattan only for work and live in other places. The most dominant function of Manhattan is a work-place where people come on weekdays during daytime. 
These regions that do not have specific functions are classified as other places with no identical temporal mobility patterns. 
The identification of such a functional region is also useful for identifying the hot area of a city at different periods.

Third, we show that these identified functional regions can help us deliver data in network applications, such as urban Delay Tolerant Networks (DTNs), more efficiently. 
DTNs \cite{dtnoffload11, TMC14,Cosense13, PhdThesis15} provide intermittent communication for humans
with mobile devices (vehicles, mobile phones, etc.), by exchanging data through short-range communications such as Bluetooth or WiFi direct. DTNs can reduce the mobile data traffic of a cellular network significantly. 

The data transferred in DTNs is delay tolerant, such as weather forecasts or sports news.  Urban DTNs provide 
complementary caching and offloading 
abilities for the congested cellular networks in a city. They can also 
provide the basic network support during disasters such as earthquakes or sudden power 
cuts in big cities. 
On the other hands, urban DTNs also benefit from the huge number of people living inside a city, e.g., tens of thousands of people gather together for a big football match. 
This high people density can help to make routing in urban DTNs more efficient.

Since humans carry their mobile devices everywhere everyday, 
understanding and utilizing urban human mobility can help in delivering the data in DTNs more efficiently \cite{TMC14}. 
The mobility patterns of different functional regions are quite different. 
Thus there are always some hot-areas inside the city during different times. We utilize this new finding and provide a simple functional-region-based urban routing algorithm. 
We show that our new algorithm has up to 183\% improvement in terms of delivery ratio 
compared to a random method. Here delivery ratio is the average ratio of the number of 
messages that successfully delivered to destinations to the total number of messages. 

The contribution of this paper is threefold. First, we find that there is a high correlation between the road networks and taxi visits. This is very 
important for many urban functional region identification applications. For example, in the previous research \cite{regionfunction} the authors require the road network information to divide the city into 
regions, then use the taxi visits to these regions as a baseline for further analysis. With this new finding we can divide the city into regions without the context information of the road network. 
Second, we provide an association-rule-based method for automatically detecting the functions (mobility patterns) of the sub-regions inside the city. 
Third, we leverage the functions of regions and urban mobility pattern for enhancing urban DTN routing. Such work can improve all other routing algorithms in the DTNs with an additional location information. 

This paper is organized as follows. Section II discusses
datasets and analysis methods. Section III presents our new
findings that there is a high correlation between road networks and 
taxi visits inside the city, and provides our new quad-tree region division methods. Section IV shows how we use the association rules for detecting the region functions.
Section V mainly discusses the potential mobile networking application
usage scenario with a DTN application example. Section VII concludes the
paper.

\section{Overview}

\subsection{Datasets}

We use three large taxi GPS trajectory datasets in our work, the Rome dataset, the San Francisco dataset and the Beijing dataset. We summarize  the key information in these three datasets in Table \ref{table:dataset}. 
All of the three datasets contain the following information: taxi id, timestamp and position (longitude, latitude).
In the taxi mobility patterns, the drivers typically either move to pick up or drop off customers, or stay in parking areas while waiting for new customers.

The San Francisco dataset \cite{SanF} is a public dataset from the Exploratorium that aims to study the invisible economic, social, 
and cultural trends of the city. The dataset contains extensive GPS data of five hundred Yellow Cab vehicles in the San Francisco region over one month (from 17th May 2008 to 10th June 2008). This dataset contains 11 million data points and the
corresponding timestamps.

The Rome dataset \cite{Rome} is a public dataset containing mobility traces of 316 taxi cabs in Rome over 30 days. 
Each taxi driver had a tablet that was set to retrieve the GPS position every 7 seconds after which the position was sent to a central server. 

The Beijing dataset \cite{Beijing1, Beijing2} is a public dataset gathered by Microsoft
Research Asia. It records the GPS trajectories of 10,357 taxis in Beijing from Feb.2 to Feb.8, 2008. There are about 15 million GPS points in this data set, and the total distance for each trajectory reaches up to 9 million kilometers.

\begin{table}[h]
    \begin{center}
    \begin{tabular}{ | l | l | l | l |}
    \hline
 & Rome & San Francisco & Beijing \\ \hline
Measurement        & GPS                   & GPS  & GPS \\ \hline                        Number of samples      &  11,219,955 & 21,817,851& 17,586,065\\ \hline
Duration               &  1 month &   1 month   &    1 week \\ \hline
Sampling Interval      & 64 s &    9 s        &    177 s\\ \hline
Number of taxis &        536  & 316    & 10357    \\ \hline
 \end{tabular}
 \end{center}
 \mbox{}
 \caption{Taxi mobility datasets of three cities}
\label{table:dataset}
\end{table}

\subsection{Obtaining Taxi Trips}

We define and extract the following information from the three datasets: GPS trajectory ($Traj$) and a Trip ($Trip$). Since taxis carry people to different places at different times, taxi trips reflect how people move in a city.

\textbf{Trajectory.} A taxi's trajectory $Traj$ is a sequence of GPS points, $Traj = (p_0, p_1, p_2, ... p_k)$, where the time interval between consecutive GPS points does not exceeds a certain threshold $\Delta T$. 
Here $p_i = (lat_i, long_i, t_i)$, where $t_i$ is the timestamp, $lat_i$, $long_i$ are the longitude and 
latitude of the GPS point respectively, $t_i < t_{i+1}$ and $t_{i+1} - t_i < \Delta T$. Here the time threshold $\Delta T$ is 30 minutes.  

\textbf{Trip.} A taxi's trip $Trip$ are the two GPS points of the taxi from picking up the passenger until dropping off the 
passenger. We believe that the trip represents the actual human movement inside the city. 
The taxi usually stops in a place (stop point) while picking up the passenger and dropping them off. 
We define the stop point $SP$ as a geographic region where the taxi stayed over a time threshold $T_{threshold}$ and within a distance threshold $D_{threshold}$. 
A taxi's trajectory $Traj$ consists of several trips $Trip$. 
Here $Trip = (p_0, p_n)$, where for all the GPS points $p_0<p_i<p_n$, the time interval between $(p_0, p_n)$ is larger than $T_{threshold}$, the distance between $(p_0, p_i)$ is smaller than $D_{threshold}$ and the distance 
between $(p_0, p_{n+1})$ is larger than $D_{threshold}$. 
We set $D_{threshold}$ as 50 meters and $T_{threshold}$ as 6 minutes.

\subsection{Basic Statistical Analysis}

Since the trips of the taxis are actually the trips of the human movement, we perform a basic statistical analysis of the urban human mobility. 
Take Rome as an example, we can obtain the trip length (how far do people move inside Rome in one trip), the trip duration (how long does it take for one trip in Rome) and the stay time (how long does the taxi stay in the same place waiting for 
the next passenger, especially important for DTN applications).  

\begin{figure*}[h]
\centering
\subfigure[Trip Length]{
\includegraphics[width=0.23\textheight]{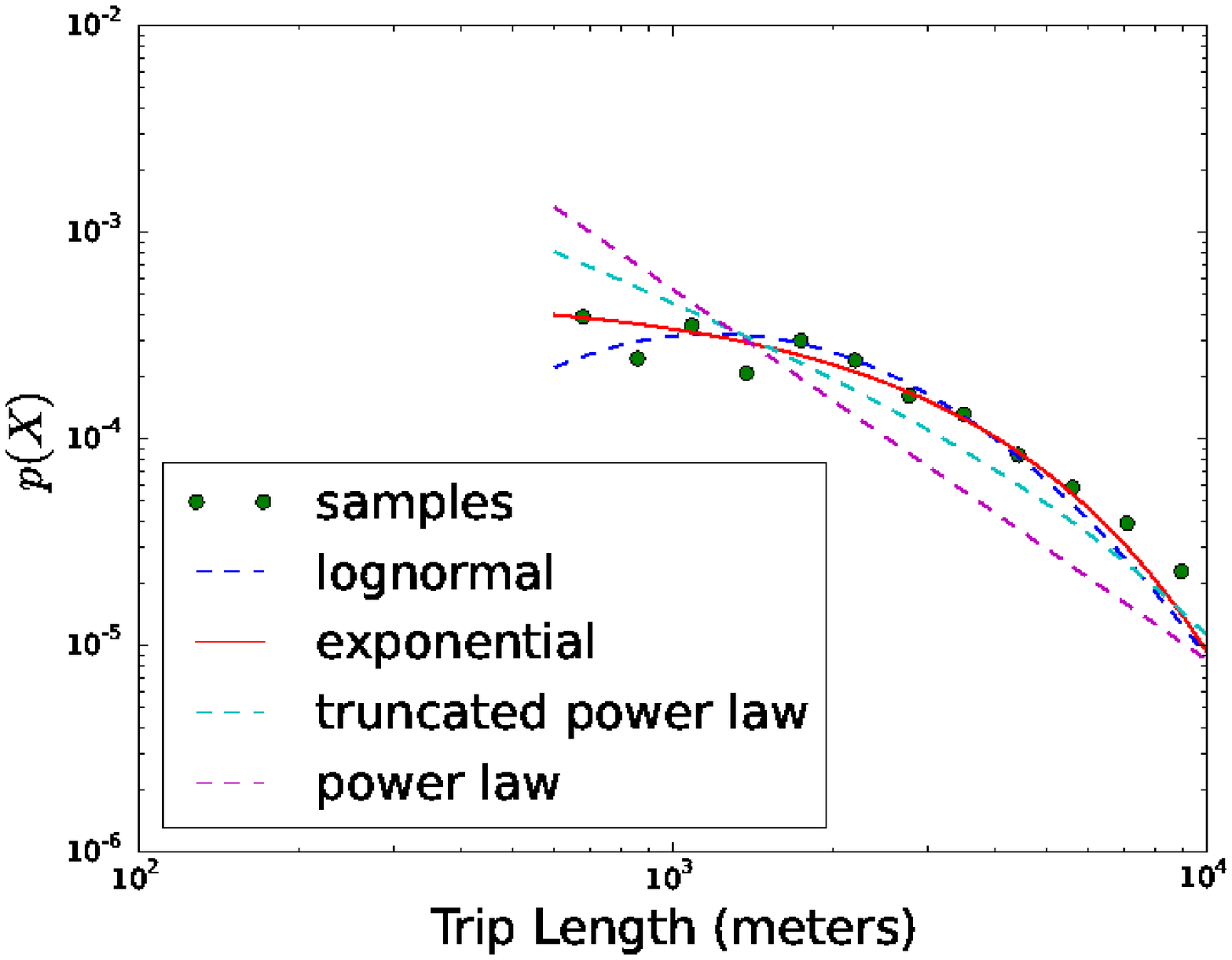}
}
\subfigure[Trip Duration]{
\includegraphics[width=0.23\textheight]{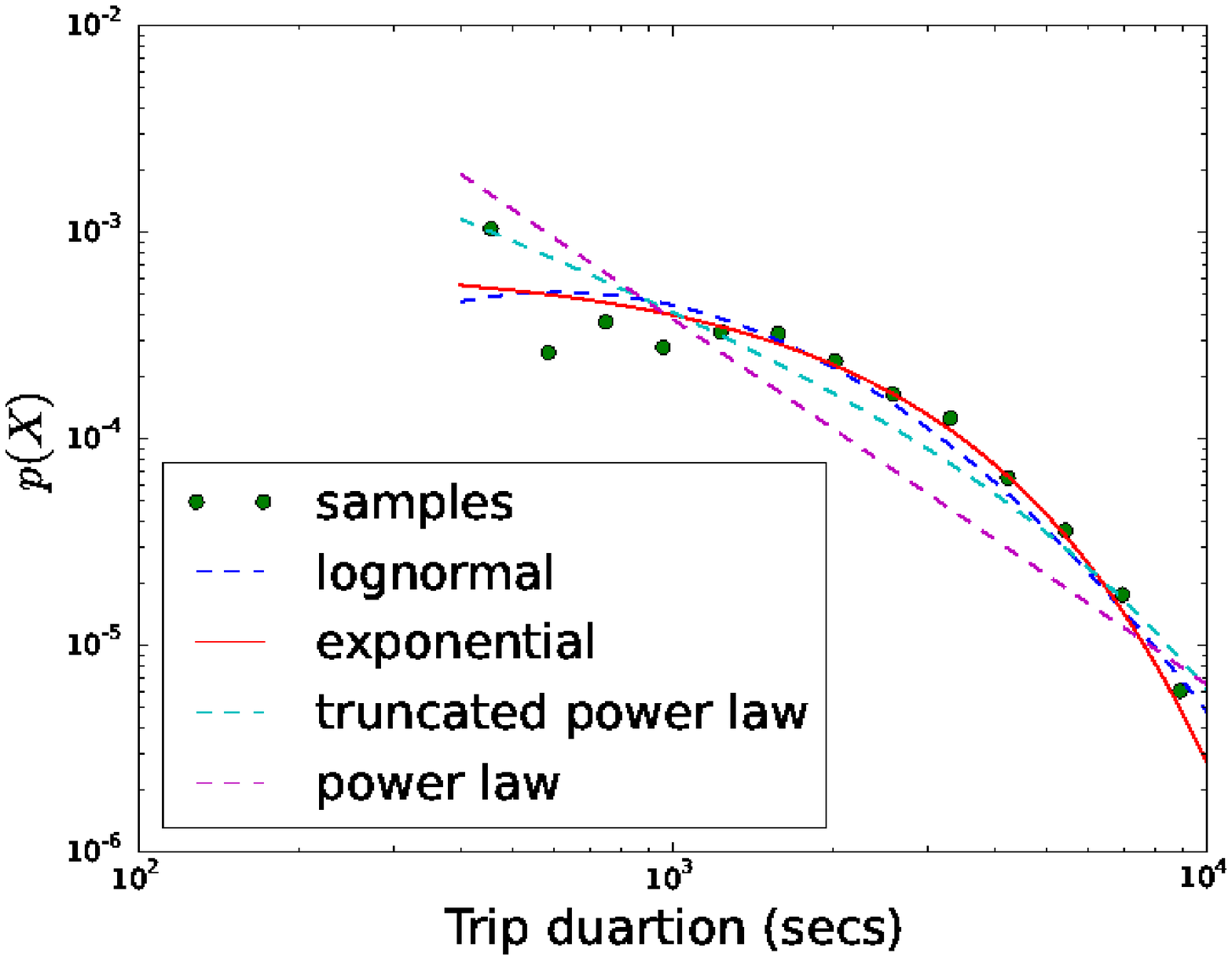}
}
\subfigure[Stay Time]{
\includegraphics[width=0.23\textheight]{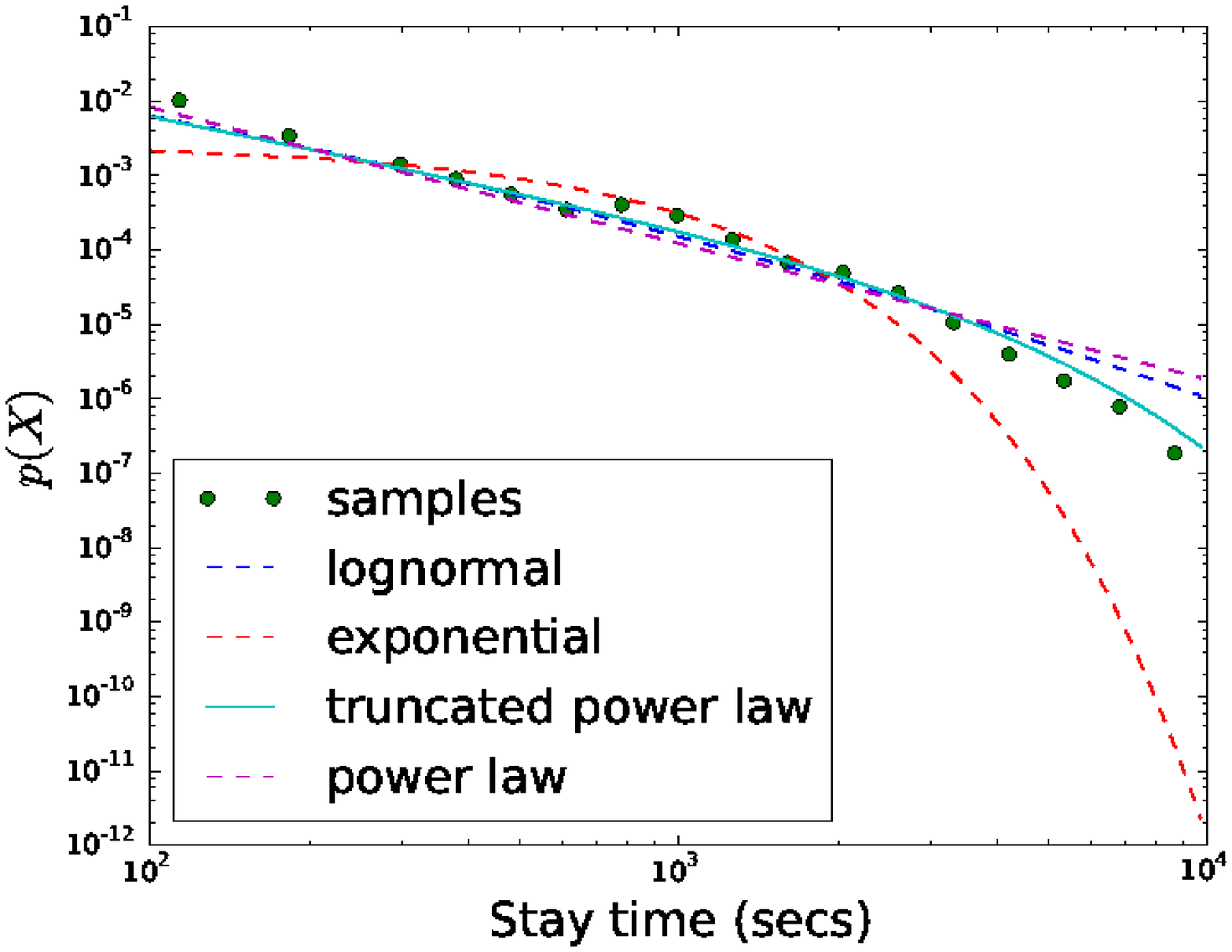}
}
\caption{Exponential distribution fitting of the trip length, exponential distribution fitting of the trip duration and power-law distribution fitting of the stay duartion}\label{fig:triplength}
\end{figure*}

We fit the trip length, the trip duration and the stay time distribution with truncated
power-law, lognormal, power-law and exponential distribution using Akaike weights \cite{AIC1}. The Akaike weight is a value between 0
and 1. The larger it is, the better the distribution is fitted. 
Fig. \ref{fig:triplength} shows fitted distributions for trip length, trip duration of humans in the city and stay time of taxis. The green points refer to the trip length, trip duration, or stay time samples obtained from the Rome dataset, 
while the solid line represents
the best fitted distribution according to Akaike weights. 

Both the trip length distribution and trip duration distribution are well fitted with an 
exponential distribution, and average trip length of a person living in Rome is 5.5 kilometres 
and the average trip duration is around 30 minutes. 
Few people tend to travel a long distance by taxi due to economic considerations. So both 
the trip length and trip duration decay very quickly and follow exponential distribution. 
The stay time distribution is well fitted with an truncated power-law distribution, and average waiting time for a taxi picking up a passenger is around 9 minutes. 

\section{Quad-tree Region Division}

In this section we first describe our new finding that there is a high 
correlation between the taxi visits and the road network intersection densities. 
Then we show that we can divide the city into different regions only based on the taxi visits, without the background information of the city 
road networks.

\subsection{Correlation Between Road Networks and Taxi Visits}

We observed that there is a high correlation between road networks and 
taxi visits. Previous work \cite{regionfunction} has used the road networks as a baseline for dividing the regions. 
However, a road network is not always presented when dividing the cities into smaller regions. Besides, the road network granularity is not always the same due to different road network layers. For example, if we only 
consider the highway and main roads, the number of total road intersections might be tens or hundreds smaller than the total number of road intersections considering all the roads including the highway, main roads and side roads. 

Instead of using the road networks, we use the taxi visits as a baseline for dividing the city into smaller regions. 
Take Beijing as an example, we use a road network dataset of Beijing containing 433,391 roads with 171,504 conjunctions and the Beijing taxi dataset containing 10,357 taxis trajectories with 17,586,065 GPS samples. We divide the entire Beijing City into 100 or 10,000 regions. 
We take the visits of the taxis in terms of entering/not
entering a specific location as a binary value. We count the number 
of intersections and the number of taxi visits falling into the same region. 
We find that the number of intersections in a region is highly correlated with the number of 
taxi visits in that region. 
We use the Pearson correlation coefficient to quantify the strength
of the correlation between them. The value of the Pearson correlation coefficient is 0.93 with a 100 region division and 0.75 with a 10,000 region division, which
means that there is a significant positive correlation between the road networks and the taxi visits inside the city.

\begin{figure}[h]
\centering
\includegraphics[width=0.43\textheight]{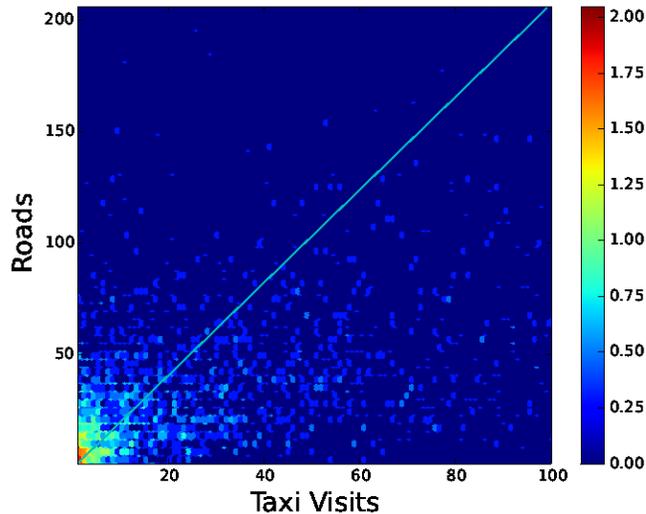}
\caption{Positive correlation of the road network intersections and taxi visits.}\label{fig:Correlation}
\end{figure}

To better illustrate the correlation between the road networks and the taxi visits, we plot the number of intersections and the number of taxi visits in the same region in Fig. \ref{fig:Correlation}. Fig. 
\ref{fig:Correlation} shows the density of the number of intersections and the number of taxi visits in the same regions. The samples are
posited near the diagonal line, which identifies a clear positive
correlation between road networks and taxi visits.

\subsection{Quad-tree Division Based on Taxi Visits}

\begin{figure*}
\centering
\subfigure[Rome]{
\includegraphics[width=0.23\textheight]{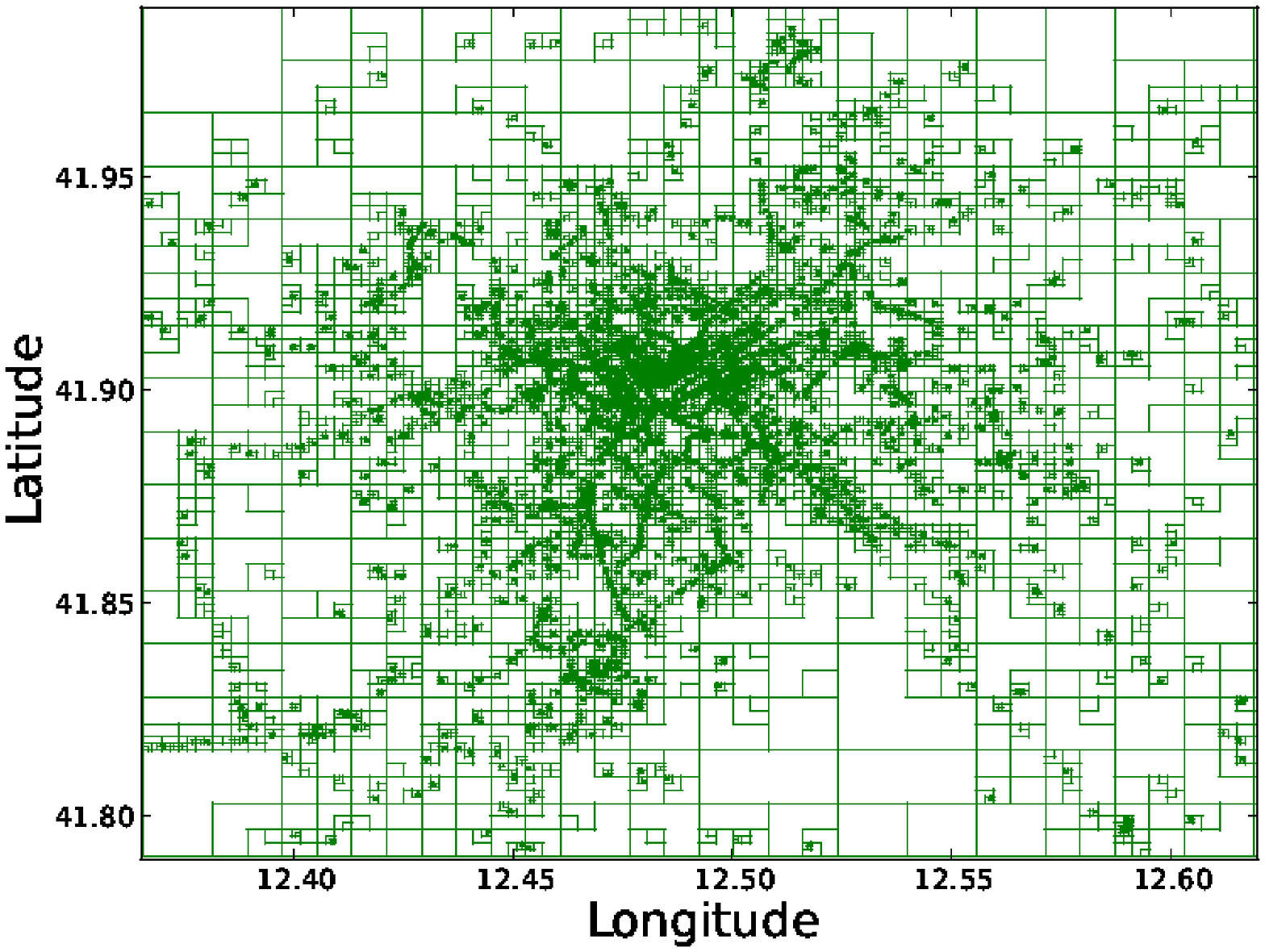}
}
\subfigure[San Francisco]{
\includegraphics[width=0.23\textheight]{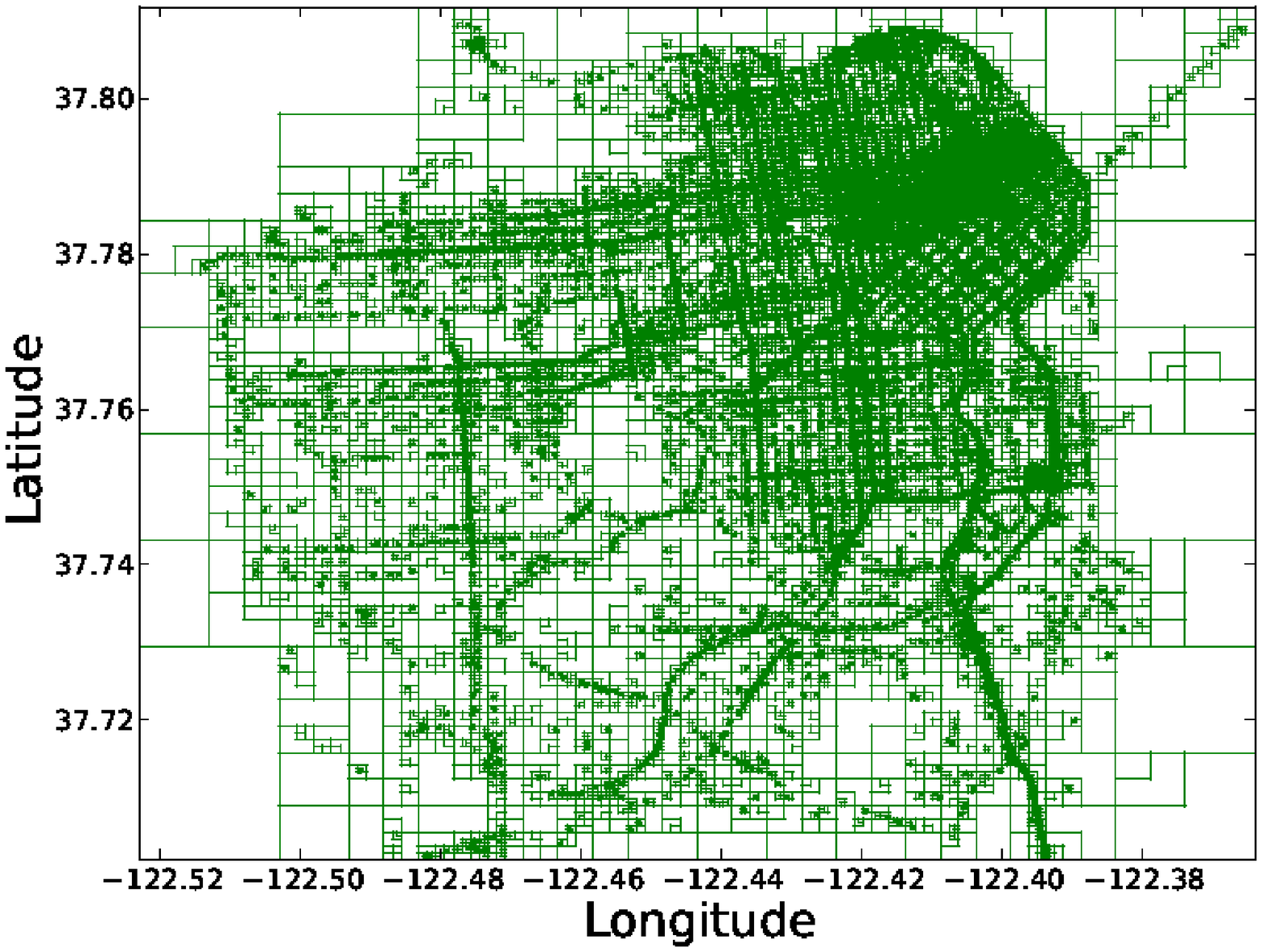}
}
\subfigure[Beijing]{
\includegraphics[width=0.23\textheight]{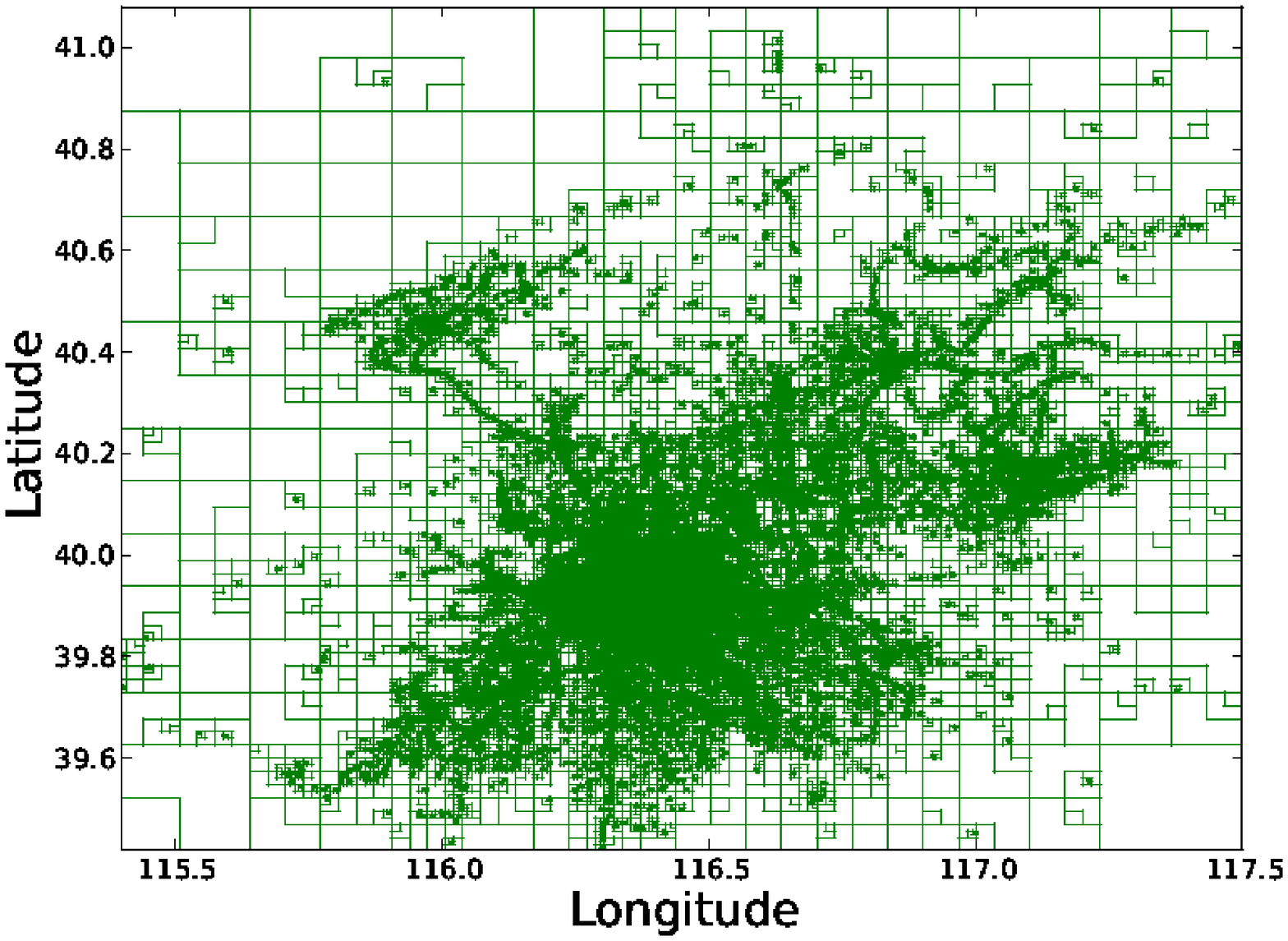}
}
\caption{Original GPS samples in three cities}\label{fig:samples}
\end{figure*}

\begin{figure*}
\centering
\subfigure[Rome]{
\includegraphics[width=0.23\textheight]{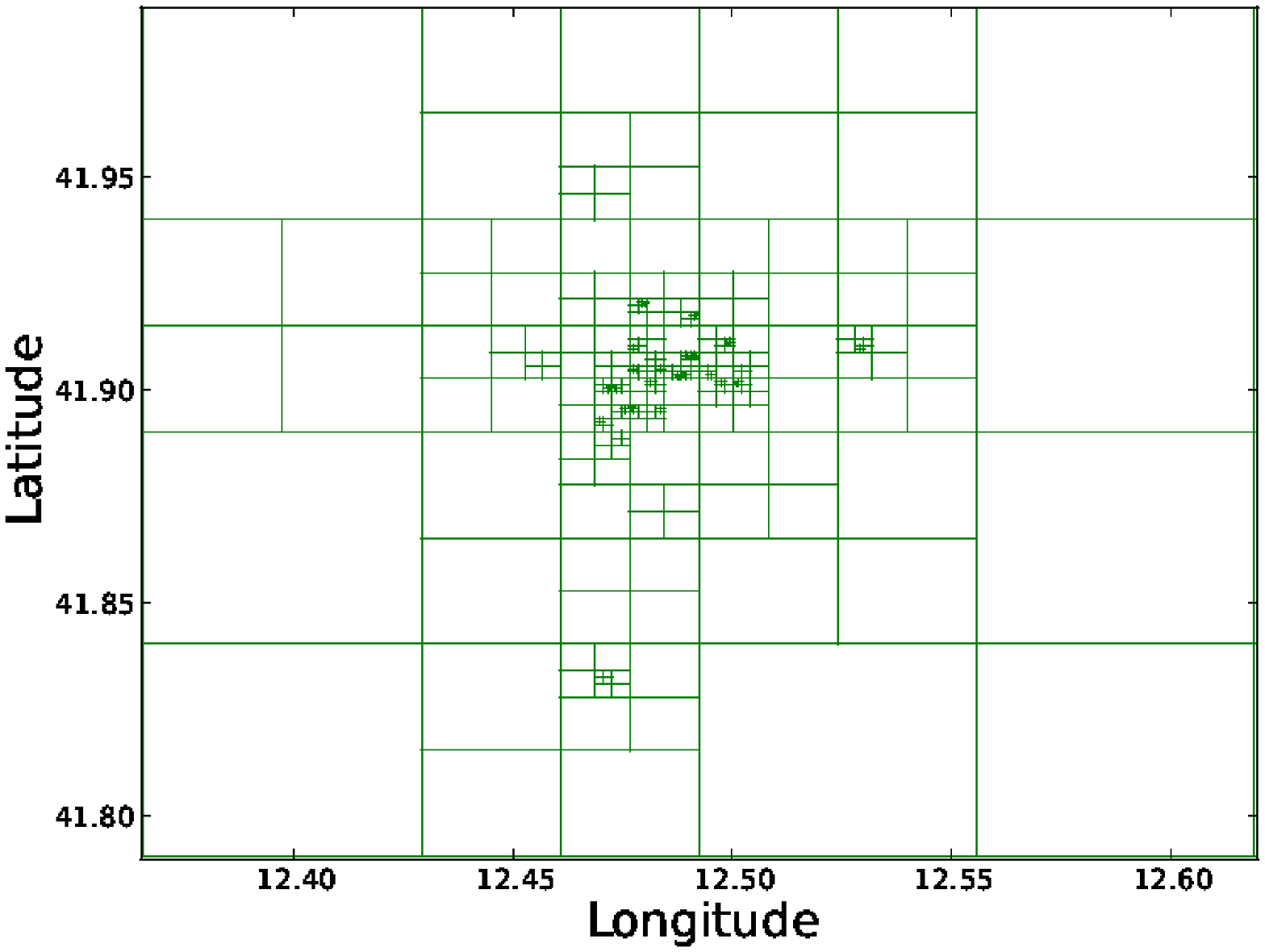}
}
\subfigure[San Francisco]{
\includegraphics[width=0.23\textheight]{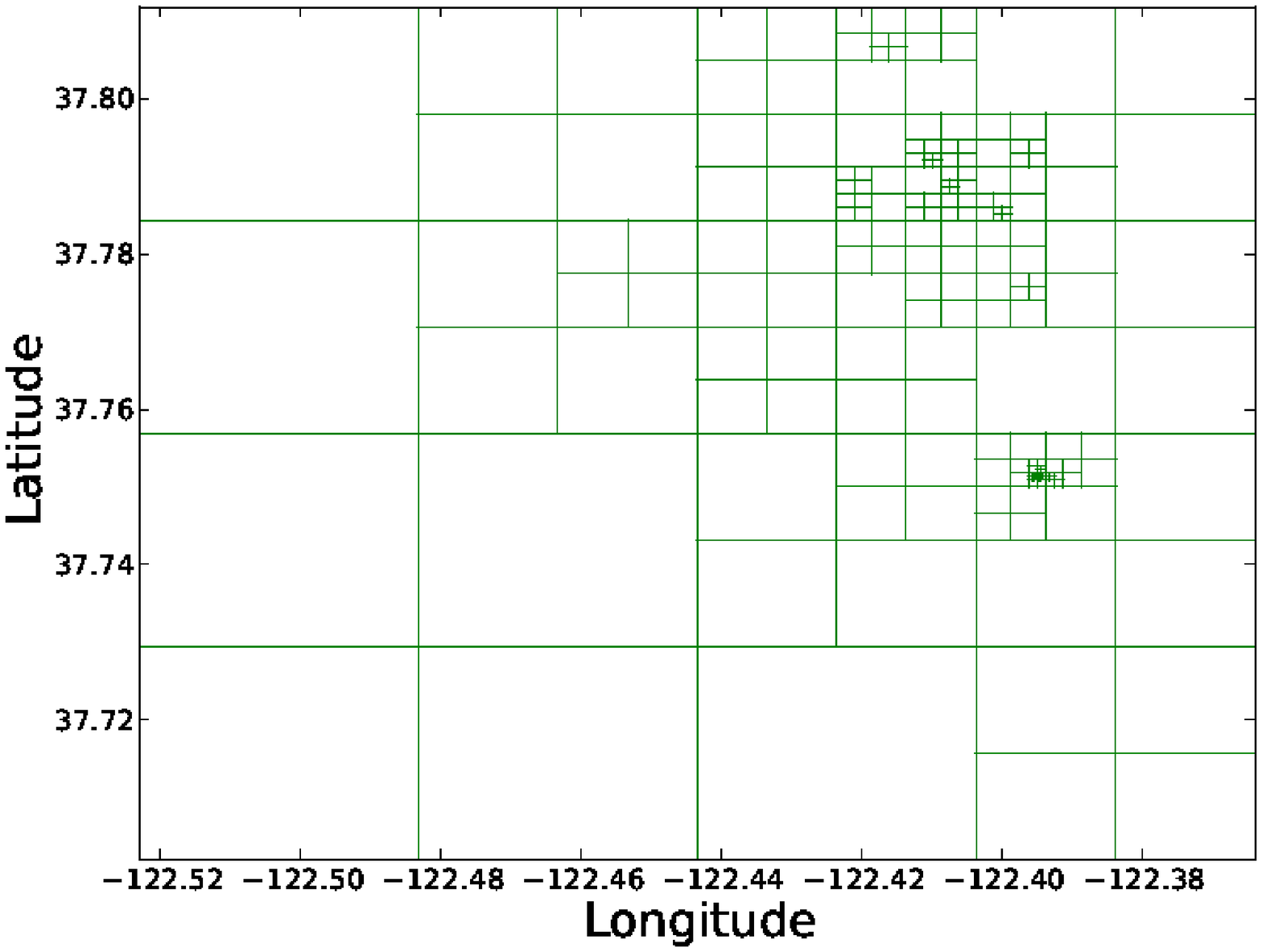}
}
\subfigure[Beijing]{
\includegraphics[width=0.23\textheight]{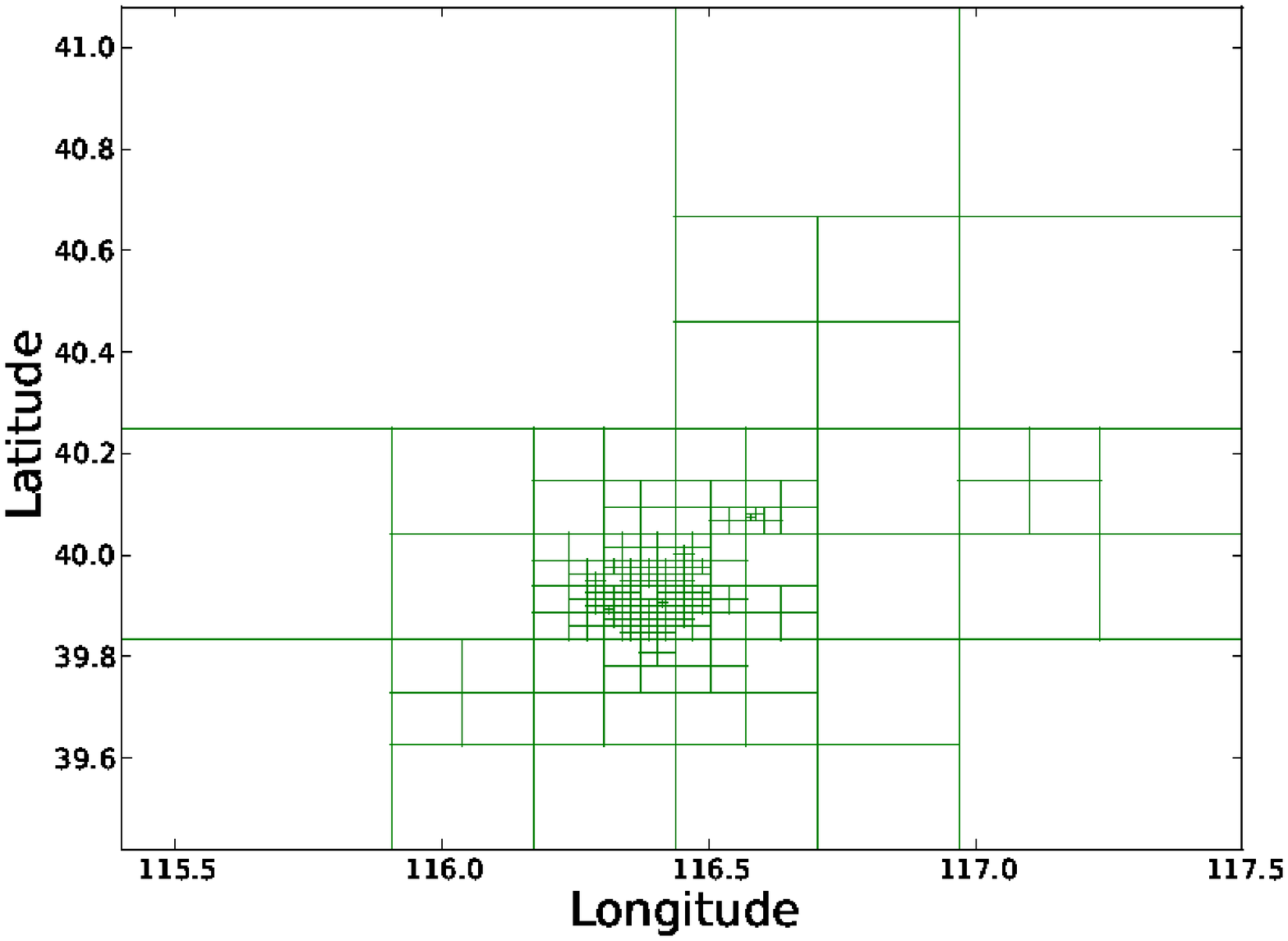}
}
\caption{Quad-tree based region division in three cities.}
\label{fig:quadtree}
\end{figure*}

Since there is a high correlation between the road networks and 
the taxi visits inside the city as shown in the previous section, we use only the taxi visits as a baseline for dividing the sub-regions. 
Fig. \ref{fig:samples} shows the GPS samples of the three cities. 
We use the quad-tree to divide the city into different regions. 
We set the threshold as 1\% of total visits inside the cities. 
If the number of visits in a region is larger than 1\% of the total visits, 
we further divide the region into four equal-sized smaller regions. 
This process continues until all the regions have equal to or less than 1\% of total visits (see Algorithm \ref{alg:quadtree}). Fig. \ref{fig:quadtree} shows the regions after the division, we obtain 367 regions in Rome, 211 regions in San Francisco and 259 regions in Beijing. 

\begin{algorithm}
  \SetKwInOut{Input}{input}
  \SetKwInOut{Output}{output}
  \Indm  
  \Input{minimum and maximum longitude pairs $P1(lat1,lon1)$ and $P2(lat2,lon2)$
threshold $t$ of allowed GPS visits $v$ in a region}
  \Output{city regions divided into $4$-sided polygons based on density of GPS visits}
\Indp
\BlankLine
 create a root node $n$, with $P1$ and $P2$ as polygon limits\;
 find all GPS visits $v$ inside polygon $n$\;
 \If{count($v$) $>$ visits threshold $Threshold_{visits}$}{
  divide node $n$ into four equal regions\;
  create four nodes $n1$,$n2$,$n3$, and $n4$ for each of divided regions\;
  assign the four nodes $n1$,$n2$,$n3$,$n4$ as children to node $n$\;
   now repeat the $step\ 3$ for all the four child nodes of $n$\;
 }
 \caption{Taxi visits density based quad tree division algorithm}
 \label{alg:quadtree}
\end{algorithm}

\begin{figure*}
\centering
\subfigure[Rome]{
\includegraphics[width=0.23\textheight]{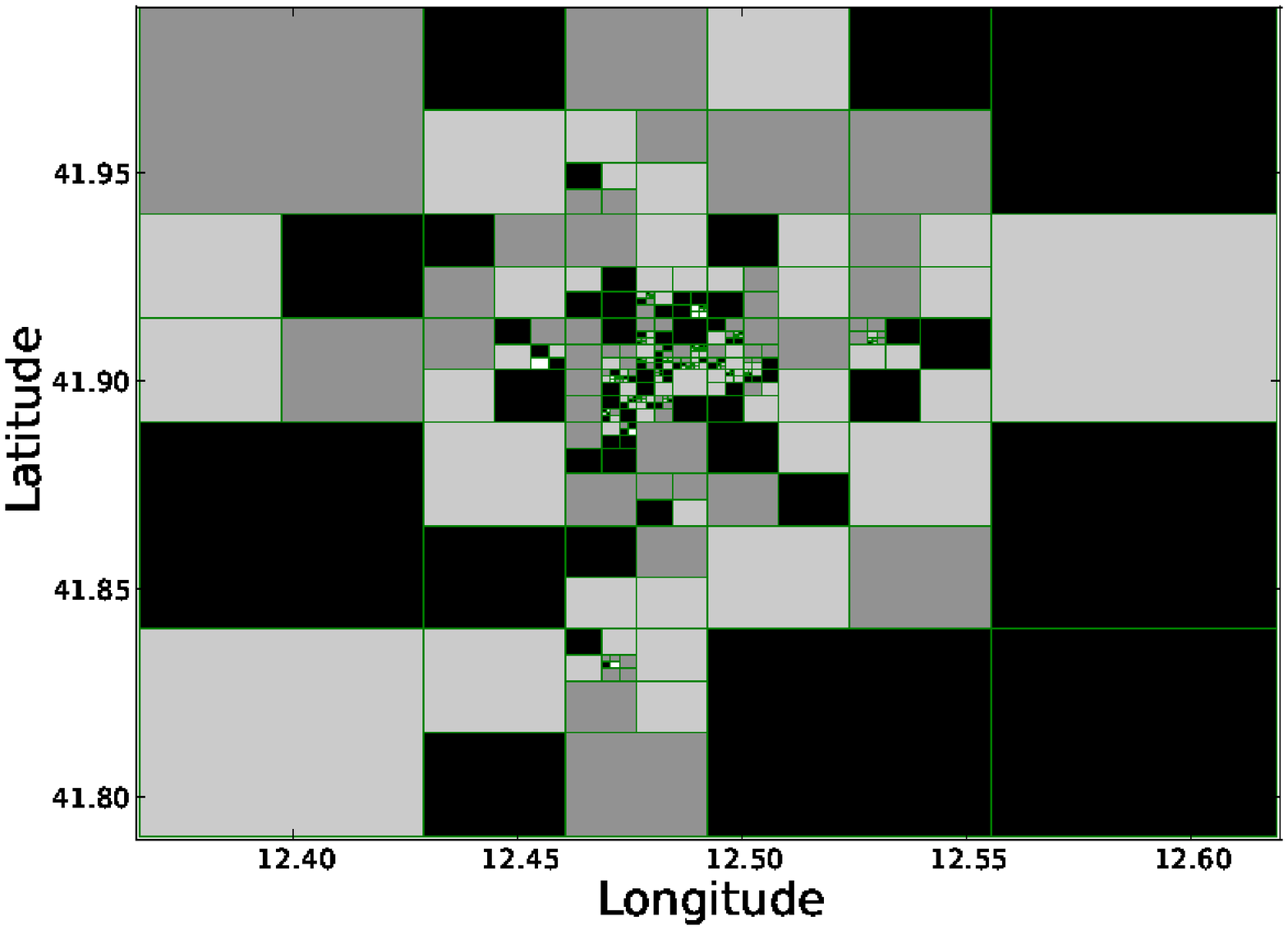}
}
\subfigure[San Francisco]{
\includegraphics[width=0.23\textheight]{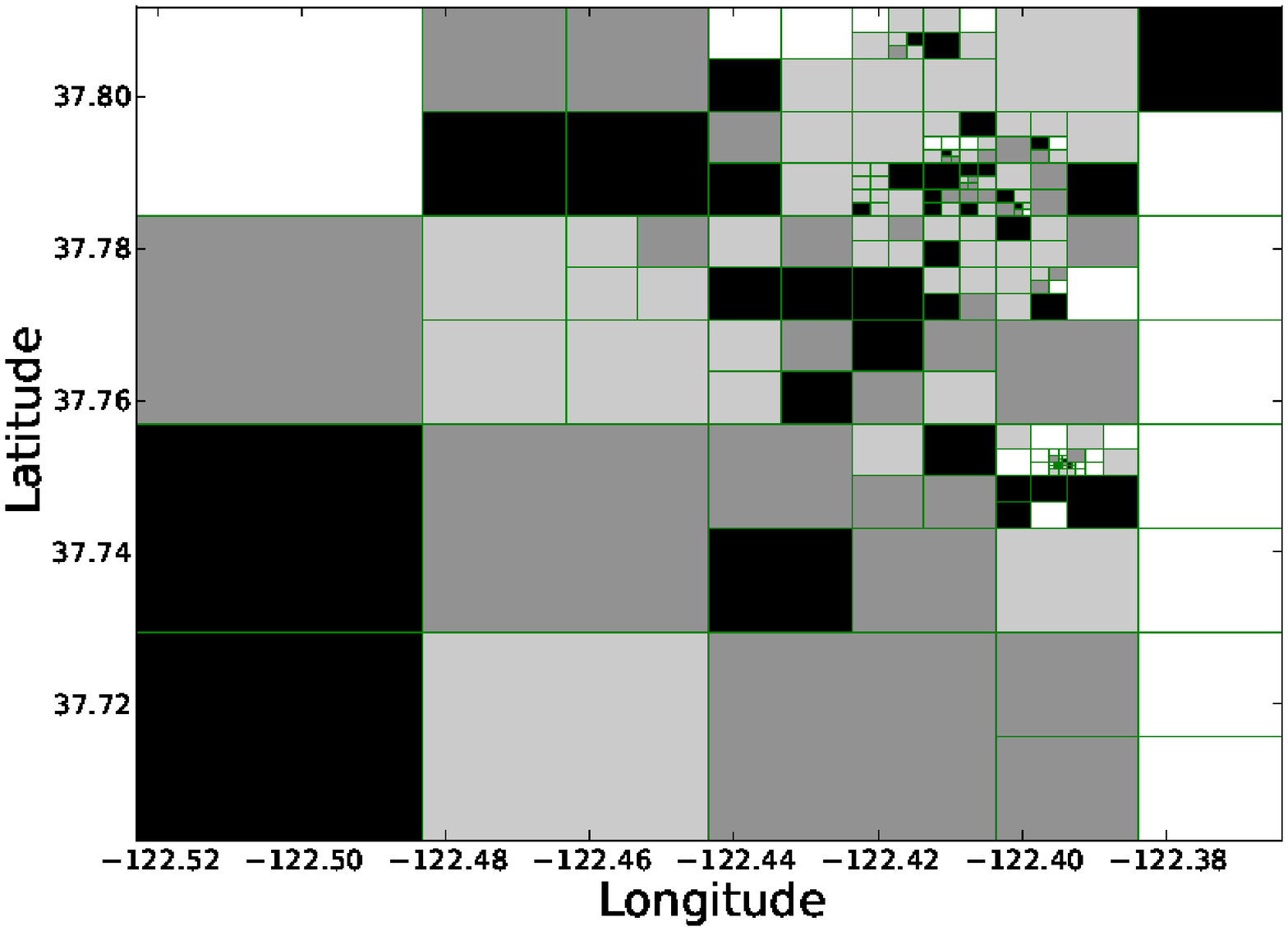}
}
\subfigure[Beijing]{
\includegraphics[width=0.23\textheight]{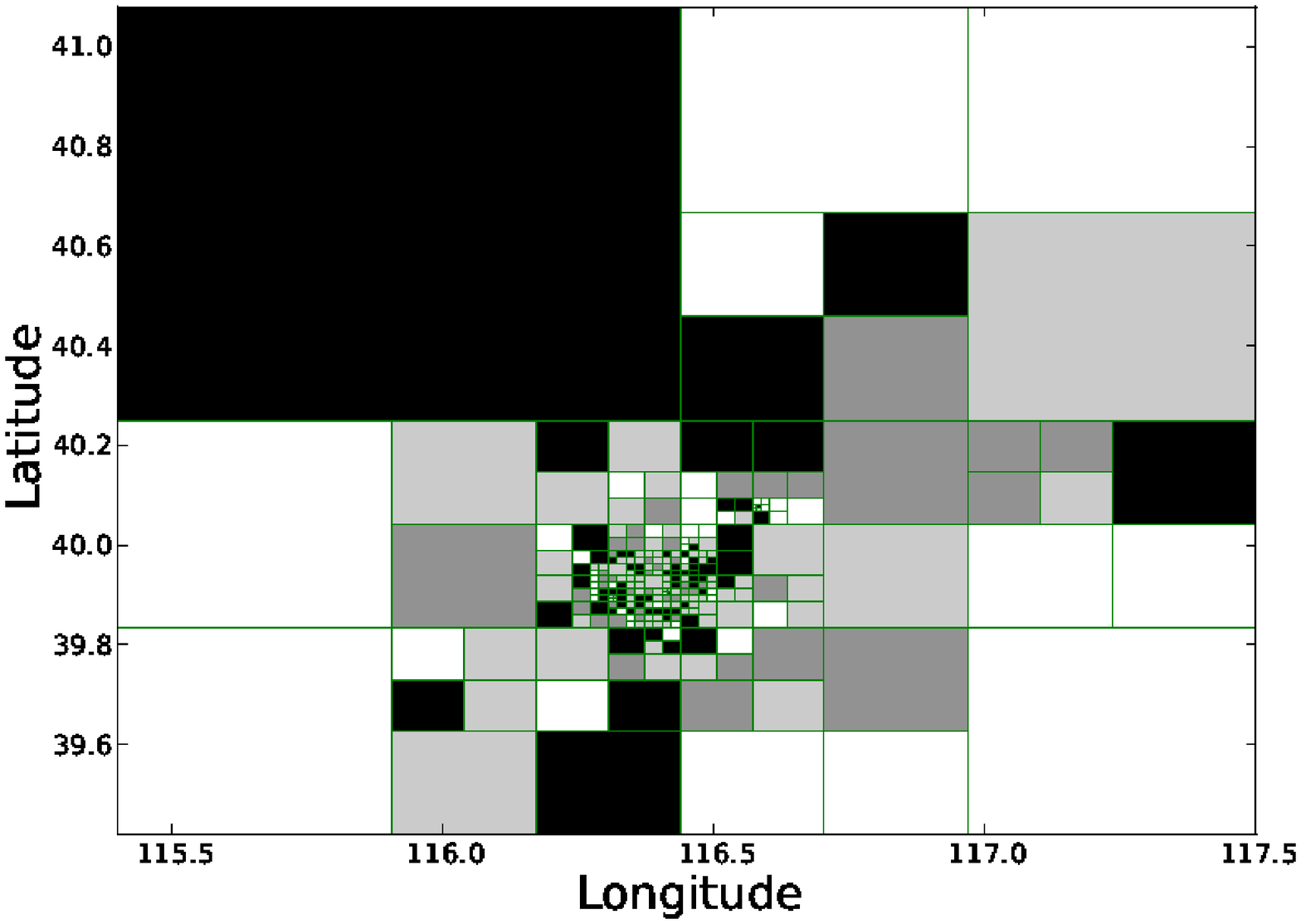}
}
\caption{Function identification in three cities. The black regions represent workplaces, 
the light grey regions represent residential places, the dark grey regions represent entertainment places while the white represent all the other places.}\label{fig:functionregions}
\end{figure*}

\section{Inferring Functional Regions}

In this section, we provide an Apriori-based \cite{DMalgorithm} function detection method for the sub-areas inside the city. 

First, we group the taxi visits for the sub-areas every hour and convert the taxi visits into a boolean table for the Apriori algorithm \cite{DMalgorithm}. 
For example, suppose there are five sub-areas and three taxis. Each taxi's visits to these five areas are {2,3,0,2,1}, {1,0,0,1,2} and {0,0,0,2,2}. 
We can build a boolean table consisting of (1,1,0,1,1),(1,0,0,1,1), (0,0,0,1,1). 
Here each transaction (row) is the list of the locations that the taxis visits. 

After building the boolean table, we use an Apriori algorithm \cite{DMalgorithm} to generate 
frequent item sets for each hour. Apriori is an algorithm for finding the frequent item sets and learning association 
rules for transactional datasets. Here in our taxi visits boolean table, 
each transaction (row) is the list of the locations that the taxi visits 
within one hour, the frequent item sets are the sets that 
people usually visit within one hour. 
We set the threshold as 0.2 to find the frequent item sets in the 
first place. 

After generating the frequent item sets (popular places people visit) 
for each hour,  we divide the city into four kinds 
of places, the workplaces which people usually visit during work time, the entertainment places where people usually stay during entertainment time, the residential places where people usually stay during home time and the other places 
with no identical mobility patterns. We define work time as the daytime on weekdays (08:00-17:00, Monday to Friday), the entertainment time as the evening on weekdays and 
daytime and evening on weekends (17:00-23:00, Monday to Friday and 08:00-22:00, Saturday to Sunday), and home time as the night time all week (23:00-08:00 Monday to Friday). 

We plot the cities according to the different 
functions detected above in Fig. \ref{fig:functionregions}. We use four colors to 
identify the functions inside the city. The black regions represent the workplaces, 
the light grey regions represent the residential places, the dark grey regions represent the entertainment places while the white represent all other places.

\section{Network Application: DTNs in the Regions}

We can utilize urban mobility patterns in different functional regions and across regions for enhancing DTNs routing algorithms. DTNs \cite{dtnoffload11, TMC14} provide intermittent communication for humans
with mobile devices (vehicles, mobile phones, etc.), by exchanging data through short-range communications such as Bluetooth or WiFi direct. 
Since humans carry their mobile devices everywhere everyday, 
understanding and utilizing human mobility in functional regions can help in delivering the data in DTNs more efficiently \cite{TMC14}. 

The key idea is to map content to functional sub-areas and then select carriers that optimize the information delivery to the destinations. For example, a person who is going to visit functional-areas during a peak time (for example, workplaces during daytime on weekdays) has a high chance of meeting a person who may need the carried content targeted for those areas. The functional regions  can also be viewed as hubs for forwarding messages to more people. 
We provide two simple functional-region-based urban routing algorithms and we show that they achieve up to 183\% improvement in terms
of delivery ratio compared with a random-routing algorithm. 

\subsection{Target-set Problem}

DTN routing protocols aim to route data (e.g., a weather forecast notification) 
from the data sources to the destinations through opportunistic and delay tolerant 
data exchanges that typically are based on short-range wireless communications, such as Bluetooth.

In DTNs, the data is not typically distributed to all devices, but only to a subset of the devices. The devices in the 
subset then further distribute the data to other devices through an opportunistic DTN communication. How we can choose the initial 
target-set for maximizing the number of devices that further receive the desired data is called the target-set problem in DTNs \cite{dtnoffload11}. 

\subsection{Oracle-based, History-based, and Random Algorithms}

We present our two simple functional region-based algorithms called the Oracle-based (Greedy) algorithm and the History-based (Heuristic) algorithm for solving the target-set problem in DTNs. To evaluate the performance of the Oracle-based and History-based 
algorithms, we compare them with the Random algorithm. In the Random algorithm, the initial group of carriers is chosen randomly.

In the Oracle-based algorithm, the initial subset of data carriers consists of 
people who have the highest probability of visiting the hot-areas in the city. 
Note that here the hot-areas represent workplaces during work time, entertainment places during entertainment time and residential places during night (see Section IV). 
The Oracle-based algorithm provides an upper bound for the target-set problem.

The History-based algorithm is similar to the Oracle-based algorithm, with the exception that a taxi’s probability of visiting hot areas is obtained from the historical data traces. We use the taxi visit data of the same hour in the previous day as the historical data. 
This strategy is motivated by the observation that human mobility has regularity. For example, a person usually goes to workplaces at 8 am in the morning, and returns for dinner time.  If the person has visited the workplaces at 8 am in the previous weekdays, he has a high probability of visiting 
the workplaces at 8 am on the current weekday.  The person would be a good candidate carrier for data to the workplaces.

\subsection{Evaluation of Functional Region-based Algorithms}

We use the Beijing dataset for evaluating the three algorithms, similar results have been 
found in the other two datasets. 
When two taxis visit (enter) the same region in Beijing at the same time, we consider it as an encounter event and these two taxis can exchange the data they carried. 
We randomly choose 100 taxis (100 subscribers) out of the 10,357 taxis as the subscribers who wish to receive messages (e.g., weather notifications) from the DTN. Then 
100 copies of the message to be delivered are distributed to 100 taxis (100 publishers). If these 100 messages are all delivered to the subscribers, we count the delivery ratio as 100\%. The initial target-set of the 100 publishers are chosen according to 
the three algorithms, the Oracle-based algorithm, the 
History-based algorithm and the Random algorithm.

After obtaining the message, each publisher will keep the message until it meets a 
subscriber and forwards the message. 
We extract the taxi visits of two hours to evaluate our algorithm under different scenarios. 
One is during "2008-02-03 (Sunday) 15:00:00 - 16:00:00" and the other one is "2008-02-05 (Tuesday) 15:00:00-16:00:00". 

During "2008-02-03 (Sunday) 15:00:00-16:00:00" these 
entertainment places are usually the hot-areas in the terms of people visits, while during "2008-02-05 (Tuesday) 15:00:00-16:00:00" these 
workplaces are usually the hot-areas. 
To obtain the historical information for our History-based algorithm, we take the taxi visits from "2008-02-02 (Saturday) 15:00:00 - 16:00:00" as a baseline for entertainment places and "2008-02-04 (Monday) 15:00:00 - 16:00:00" as 
a baseline for workplaces. 

\begin{figure}[h]
\centering
\includegraphics[width=0.37\textheight]{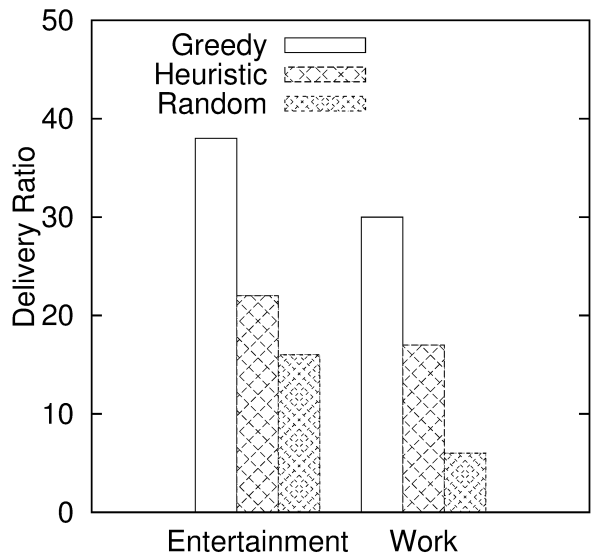}
\caption{Delivery ratio of a location-based Oracle-based (Greedy) algorithm, History-based (Heuristic) algorithm and Random algorithm.}\label{fig:ratio}
\end{figure}

Fig. \ref{fig:ratio} shows the delivery ratio of the three algorithms. The delivery ratio is the average ratio of the number of data messages that were successfully
delivered divided by the total number of messages. 
We observe that the Oracle-based algorithm outperforms the other two as it provides the upper bound. The History-based algorithm increases the deliver ratio with up to 183\% improvement compared with a Random algorithm. This is mainly because these taxis carrying data (e.g., today's weather forecast) visiting the hot-areas (e.g., workplaces during daytime on weekdays) have a higher chance of meeting a person 
who needs the required data (e.g., today's weather forecast). These functional regions can be viewed as hubs for 
forwarding messages to more people.

\section{Conclusion}

In this paper we infer the functions of regions in three cities such as workplaces, entertainment places, residential places. 
First, we provide a new quad-tree region division algorithm based on the taxi visits. 
Then we use association rules to infer the functions of different regions in these three cities. Finally we show that these identified regions can help us to deliver data in DTNs with up to 183\% of improvement in terms of delivery ratio. 

Since this is still an ongoing work, some important parts are still missing. Adding Points of
Interest (POIs) will further increase the accuracy of the region function identification. A comparison with other traditional transportation engineering works is needed for the evaluation of our method. The computational improvement of the quad-tree method has not yet been evaluated. We will consider these in future work. 

\section*{Acknowledgment}

The work was supported in part by the Digile IoT SHOK program and the Tekes ESENS research project.

\bibliography{globecomm2}
\bibliographystyle{abbrv}

\newpage

\section*{Appendix. a Akaike weights}

We use Akaike weights to choose the best fitted distribution. An Akaike weight is a normalized distribution selection criterion.
Its value is between 0 and 1. The larger the value is, the better the distribution is fitted.  

Akaike's information criterion (AIC) is used in combination with Maximum likelihood estimation (MLE). 
MLE finds an estimator of $\hat{\theta}$ that maximizes the likelihood function $L(\hat{\theta}|data)$ 
of one distribution. 
AIC is used to describe the best fitting one among all fitted distributions, 
\begin{equation} 
AIC = -2 log \left(L(\hat{\theta}|data)\right) + 2K.
\end{equation}
Here K is the number of estimable parameters in the approximating model.

After determining the AIC value of each fitted distribution, we normalize these values as follows. 
First of all, we extract the difference between different AIC values called $\Delta_i$, 
\begin{equation}
\Delta_i = AIC_i - AIC_{min}. 
\end{equation}

Then Akaike weights $W_i$ are calculated as follows,
\begin{equation} 
W_i = \frac{exp(-\Delta_i / 2)}{\sum_{r = 1}^{R} exp(-\Delta_i / 2)}.
\end{equation} 

\section*{Appendix. b Apriori algorithm} 

\begin{algorithm}
\Indp
 $k$ = 1\;
 $F_{k}\ =\ \{\ i\ |\ i\ \in\ I \land \sigma\{i\}\ \geq\ N\ \ast \ minsup  \}$\{Find all frequent 1-itemsets\}\;
 \Repeat{}{ 
    $k$ = $k$ + 1\;
    $C_k\ =\ apriori-gen(F_{k-1}).$\ \ \ \{Generate Candidate itemsets\}\;
    \For{$each\ transaction\ t\ \in\ T \ \do$}{
      $C_{t}\ =\ subset(C_{k},t).$\ \ \{Identify all candidates that belong to t\}\;
      \For{$each\ candidate\ itemset\ c \in C_{t} \do$}{
	  $\sigma(c)=\sigma(c)\ +\ 1$\ \ \ \{Increment support count\}\;
      }
    }
    $F_{k}=\{ c|c\in C_{k}\land\sigma(c) \geq N\ \ast\ minsup\}$\{Extract the frequent k-itemsets\}\;
  } ($F_{k}=\emptyset$)
  Result = $\cup F_{k}$\;
\caption{Frequent itemset generation of the Apriori algorithm}
 \label{alg:Apriori}
\end{algorithm}

\end{document}